\documentclass[aps,twocolumn,showpacs,superscriptaddress]{revtex4} 
\usepackage{graphicx}
\usepackage{bm}
\usepackage[hidelinks]{hyperref}
\usepackage{dcolumn,xcolor,ulem}
\usepackage{siunitx}
\usepackage{amsmath} 
\usepackage{amssymb}
\usepackage{amsfonts}
\usepackage[latin1]{inputenc}

\renewcommand{\em}{\it}

\begin{document}

\title{Power-law creep and residual stresses in a carbopol microgel}

\author{Pierre~Lidon}
\author{Louis~Villa}
\author{S\'{e}bastien~Manneville} \affiliation{Univ Lyon, Ens de Lyon, Univ Claude Bernard, CNRS, Laboratoire de Physique, F-69342 Lyon, France}

\date{\today}

\begin{abstract}
We report on the interplay between creep and residual stresses in a carbopol microgel. When a constant shear stress $\sigma$ is applied below the yield stress $\sigma_\text{y}$, the strain is shown to increase as a power law of time, $\gamma(t)=\gamma_0 + (t/\tau)^\alpha$, with an exponent $\alpha=0.39\pm 0.04$ that is strongly reminiscent of Andrade creep in hard solids. For applied shear stresses lower than some typical value $\sigma_\text{c}\simeq 0.2 \sigma_\text{y}$, the microgel experiences a more complex, anomalous creep behaviour, characterized by an initial decrease of the strain, that we attribute to the existence of residual stresses of the order of $\sigma_\text{c}$ that persist after a rest time under a zero shear rate following preshear. The influence of gel concentration on creep and residual stresses are investigated as well as possible aging effects. We discuss our results in light of previous works on colloidal glasses and other soft glassy systems. 
\end{abstract}

\maketitle

\section{Introduction}
\label{sec:intro}

In the introduction of his 1916 paper, Eugene Cook Bingham states: ``In the study of plastic flow it has already been shown that most homogeneous solids will flow somewhat after the manner of liquids, if subjected to sufficient pressure. Copper, steel, lead, ice, menthol, glass, and asphalt fall in this class insofar as they may be regarded as homogeneous solids. But ordinarily plastic substances are not homogeneous solids but suspensions of finely divided solids in fluids, such as paint in oil, lime in water, and especially clay in water. Numerous papers have been devoted to the explanation of this latter type of plasticity.'' \cite{Bingham:1916}. A hundred years later, in spite of an overwhelming number of works on the subject, the details of plastic flow in such amorphous materials as the clay suspensions studied by Bingham,   nowadays referred to as ``soft glassy materials,'' still resist complete understanding \cite{Lin:2016}.

Creep, the slow deformation that results from the application of a constant stress on a solidlike material, is a common manifestation of plasticity. Six years before Bingham's paper, at the age of 23, Edward Neville da Costa Andrade showed that when stretched under a constant load, some metallic wires deform such that the strain grows with time as a power law of exponent 1/3 \cite{Andrade:1910}. This behaviour, known as ``power-law creep,'' has been interpreted in terms of peculiar dislocation motion within the crystalline matrix due to thermally activated processes \cite{Cottrell:1952,Mott:1953,Cottrell:1997} or due to dislocation jamming \cite{Miguel:2002}. Although also observed in a variety of disordered materials including emulsions, microgels and colloidal gels and glasses, the origin of power-law creep remains mostly elusive in such soft amorphous systems \cite{Sentjabrskaja:2015}.

When the applied stress is increased above some critical value, creep gives way to fracture or to viscous flow depending on the material microstructure. In his 1916 paper, Bingham focused on ``English china clay.'' Using a capillary viscometer, he  showed that this colloidal dispersion was well described by the following affine constitutive relation between the stress $\sigma$ and the shear rate $\dot\gamma$: $\sigma=\sigma_\text{y}+\eta\dot\gamma$, where $\sigma_\text{y}$ is the yield stress of the material and $\eta$ is the viscosity of the material when made to flow far above $\sigma_\text{y}$. Ten years later, this empirical relation was generalized by Winslow Herschel and Ronald Bulkley to $\sigma=\sigma_\text{y}+k\dot\gamma^n$, where $k$ and $n$ are known as the consistency index and flow index respectively \cite{Herschel:1926}. This relation accounts for shear-thinning above the yield stress and the exponent $n$ typically lies between 0.2 and 0.8 \cite{Roberts:2001,Bonn:2015}. The Herschel-Bulkley (HB) constitutive relation was found to account correctly for the behaviour of a wide range of disordered materials such as foams, emulsions and microgels, which were recently coined ``simple yield stress fluids'' by contrast with other soft glassy materials that display more complex rheological behaviour including thixotropy or shear localization \cite{Moller:2009b,Balmforth:2014,Bonn:2015}.

An additional well-known feature of glassy systems is their history-dependence. For instance, when a silica glass is thermally quenched below the glass temperature, it stores stresses that remain trapped within the material. These ``residual'' or ``internal'' stresses are routinely used to strengthen so-called prestressed glasses in industrial applications. Similarly, soft glassy materials store residual stresses when mechanically quenched from a presheared fluidised state to rest at zero strain or stress \cite{Ballauff:2013}. The configuration reached by the quenched microstructure is generally metastable and evolves more and more slowly as the system explores more stable configurations. Such a physical aging may subsequently compete with rejuvenation induced by shear \cite{Cloitre:2000,Viasnoff:2002}.

The main goal of the present contribution is to explore creep in a carbopol dispersion, a system made of acidic cross-linked polymer particles that swell and jam upon neutralization \cite{Ketz:1988,Piau:2007,Putz:2009}. Such carbopol ``microgels'' generally nicely follow the HB behaviour and are considered as non-aging, non-thixotropic yield stress materials \cite{Curran:2002,Piau:2007,Lee:2011}. However, a number of recent results have reported unexpected results for such a simple yield stress fluid, including rheological hysteresis \cite{Putz:2009,Divoux:2013}, transient shear localization \cite{Divoux:2012}, influence of rest time on stress overshoots upon shear start-up \cite{Divoux:2011b} and influence of confinement on flow properties \cite{Geraud:2013a}. Creep below the yield stress has only been scarcely explored in carbopol and previous works focused on the debate about the existence of a true yield stress \textit{vs} a Newtonian plateau at low shear rate \cite{Moller:2009a,Dimitriou:2013}.

Here, we focus on a carbopol ETD~2050 dispersion whose preparation, microstructure and basic rheological features are discussed in Section~\ref{sec:material}.  Section~\ref{sec:results} presents the results of creep experiments performed after preshear followed by a rest time under a zero shear rate in cone-and-plate geometry. We first show that our microgels display robust power-law creep with an exponent $\alpha=0.39\pm 0.04$ for stresses ranging from the yield stress $\sigma_\text{y}$ down to a typical value $\sigma_\text{c}\simeq 0.2 \sigma_\text{y}$. For applied shear stresses lower than $\sigma_\text{c}$, the microgels experience a more complex creep behaviour, characterized by an initial decrease of the strain. We attribute this anomalous creep to residual stresses of the order of $\sigma_\text{c}$ that slowly relax after quenching the system from a presheared fluidised state and that compete with the applied stress. This behaviour appears to be independent of the carbopol concentration (Appendix~B) but is no longer observed if one prevents residual stresses by imposing a zero shear stress during the rest time (Appendix~C). In Section~\ref{sec:discussion}, these results are discussed in light of the current literature and compared to previous works on other systems in terms of glassy dynamics and aging phenomena. Beside findings specific to carbopol microgels, we emphasize the relevance of creep studies for a better general understanding of soft glassy materials and for deeper investigations of the subtle interplay between microstructure and deformation below the yield stress.

\section{Materials and methods}
\label{sec:material}

\subsection{Sample preparation, microstructure and rheology}
\label{sec:preparation}

Our samples are made of carbopol ETD~2050 following the preparation protocol described in \cite{Geraud:2013a,Geraud:2013b}. The carbopol powder, a mixture of homopolymers and copolymers of acrylic acid cross-linked with a poly\-alkyl poly\-ether, is dissolved in deionized water at 50$^\circ$C under magnetic stirring for 30~min. The carbopol mass fraction is $C=1$~\%~wt throughout this paper, except in Appendix~B where mass fractions of 0.6~\%~wt and 2~\%~wt are briefly explored. After equilibration at room temperature for 30~min, the acid solution is neutralized by addition of sodium hydroxyde (NaOH at 0.1~M) until the pH reaches a value of $7.0\pm 0.5$. This leads to a dilution of the carbopol solution by about 10~\% so that the final carbopol mass fraction is close to 0.9~\%~wt. Neutralization induces the swelling and jamming of polymer particles \cite{Roberts:2001,Baudonnet:2004}. The sample is subsequently homogenized for 24~h by stirring with a mixer (RW20, Ika, tip: R1303) at  2000~rpm. Finally, the sample is centrifuged for 10~min at 3200~rpm in order to remove air bubbles. 

The microstructure of carbopol ETD~2050 dispersions has been characterized by dynamic light scattering and confocal imaging \cite{Lee:2011,Gutowski:2012,Geraud:2013b}. It was shown that these systems can be considered as a dense, amorphous assembly of soft jammed particles made of swollen polymers, namely as a ``microgel'' following a widespread --yet rather loose-- definition \cite{Baker:1949,Saunders:1999,Roberts:2001}. The typical size of the particles is about $2~\mu$m as estimated from confocal microscopy on the same system as that investigated here \cite{Geraud:2013b}. Such a length scale is consistent with light scattering measurements performed on samples made of the same carbopol and prepared in similar conditions \cite{Lee:2011}. More specifically, \cite{Lee:2011} report the existence of two different length scales, which points to a heterogeneous structure where regions of  higher- and lower- than-average cross-link density coexist. This suggests that carbopol ETD~2050 consists of randomly cross-linked polymers rather than polymer particles with heavily cross-linked cores and dangling free ends.

As a result of the percolation of the network of swollen polymer particles, carbopol dispersions typically show a yield stress for weight concentrations as low as 0.06~\% \cite{Roberts:2001,Oppong:2006,Oppong:2011}. For concentrations  larger than typically 0.2~\%~wt, rheological measurements suggest that the system can be described as a ``space-filling paste of swollen microgel particles'' \cite{Gutowski:2012}. As shown in Appendix~A (see Fig.~\ref{fig:flowcurve}), the flow curve of our carbopol ETD~2050 microgel for $C=1$~\%~wt nicely follows HB rheology, $\sigma=\sigma_\text{y}+k\dot{\gamma}^n$, with parameters $\sigma_\text{y} =10.2$~Pa, $n=0.60$ and $k=3.3$~Pa.s$^n$ that are fully consistent with those reported on the same system with similar preparation protocols \cite{Divoux:2010,Divoux:2012,Gutowski:2012,Geraud:2013a}. The large extent of the linear regime up to about one strain unit, the low-frequency values of the viscoelastic moduli $G'\simeq 37$~Pa $\gg G''\simeq 3$~Pa and their weak increase with frequency (see Fig.~\ref{fig:oscill} in Appendix~A) are also in line with previous measurements available in the carbopol literature \cite{Ketz:1988,Benmouffok:2010,Divoux:2011b,Gutowski:2012,Geraud:2013b,Jofore:2015}.
These various rheological features are characteristic of ``soft glassy materials'' \cite{Liu:1996b,Fielding:2000,Derec:2003}. Although one should keep in mind that there is no universality among these systems due to the huge variety of microstructures, this ``simple'' yield stress rheology makes carbopol microgels good candidates for generic studies of creep phenomena.

\subsection{Creep test and flow cessation protocols}
\label{sec:creepprotocol}

All the experiments reported here were performed with a stress-imposed rheometer (Anton Paar MCR 301) equipped with a cone-and-plate geometry of diameter 40~mm and angle $2^\circ$. To minimize wall slip, we used a sand-blasted cone (CP40-2/S) with a roughness of 5.5~$\mu$m. The bottom plate includes a Peltier element that keeps the temperature fixed to 25$^\circ$C. A specific cover is used to limit evaporation and allows us to work on the same sample loading for at most $\sim 12$~h.

In Sect.~\ref{sec:creep} below, we perform creep experiments by monitoring the strain response $\gamma(t)$ to a constant shear stress $\sigma$. As in most soft glassy materials, a good control of the initial state of the microgel sample is crucial to warrant reproducibility and to ensure proper analysis of the experimental data. Therefore, prior to any experiment (including the previous viscoelastic measurements), the freshly loaded sample is first thoroughly fluidised thanks to a strong preshear under a controlled shear rate $\dot{\gamma}_\text{p}$ for a duration $t_\text{p}$. The sample is then allowed to rest under a zero shear rate during a waiting time $t_w$. Such a protocol allows us to start subsequent tests in a reproducible initial state. We note that it slightly differs from usual creep protocols performed from a rest period under a zero shear stress \cite{Cloitre:2000,Purnomo:2007}. Experiments performed by imposing a zero shear stress rather than a zero shear rate during the rest time are briefly presented in Appendix~C. We check for reproducibility (and for lack of evaporation or other long-term degradation of the sample) by systematically measuring the viscoelastic moduli during the rest time under small-amplitude oscillatory shear (strain amplitude 1~\% and frequency 1~Hz). The variation in $G'$ from one experiment to the other over the same loading is found to be less than 1~\%. 

Unless otherwise specified, we use $\dot{\gamma}_\text{p}=\SI{100}{\per\second}$, $t_\text{p}=\SI{60}{\second}$ and $t_\text{w}=\SI{300}{\second}$ as default parameters for the preshear and rest protocol. The creep test is started immediately after the rest time and the strain $\gamma(t)$ is monitored for at least 300~s and up to one hour. Thus, the origin of time $t$ is taken here after an ``aging'' duration $t_w$ following preshear. The influence of $\dot{\gamma}_\text{p}$ and $t_\text{p}$ will be addressed in Sect.~\ref{sec:interplay}. Aging issues, i.e. a possible dependence with $t_w$, will be discussed in Sect.~\ref{sec:aging}. 

We checked that a constant torque is reached within about 10~ms and that it is kept perfectly constant by the rheometer throughout the creep experiment. However, inertio-elastic oscillations occur upon application of stress and the material is truly submitted to a constant stress only once these oscillations are damped, typically after a couple of seconds (see Fig.~\ref{fig:creep_fit} and discussion below). After a creep test under a given shear stress $\sigma$, the preshear and rest protocol is repeated over again and another creep test is performed for a different value of $\sigma$. Successive values of $\sigma$ are chosen in a random order to avoid any artefact arising from a systematic drift in the sample or in the setup. This procedure is iterated until the elastic modulus at rest differs by more than 10~\% from its initial value, indicating evaporation and/or sample alteration, thus requiring to load a new sample. To improve the statistics, creep experiments are reproduced three to six times for the same shear stress on different loadings of the same microgel preparation batch.

Finally, in Sect.~\ref{sec:interplay}, we shall investigate stress relaxation after preshear through flow cessation experiments. The reason for these additional experiments is that our rheometer does not provide the full temporal stress response under small-amplitude oscillatory strain but only the amplitude of the stress oscillations, so that stress relaxation cannot be monitored simultaneously to the viscoelastic moduli. The preshear protocol for flow cessation experiments is the same as for creep experiments, namely preshear under a controlled shear rate $\dot{\gamma}_\text{p}$ for a duration $t_\text{p}$. After preshear, a zero shear rate is applied and the stress $\sigma(t)$ is monitored over time $t$ whose origin is taken immediately after preshear. 

\section{Results}
\label{sec:results}

\subsection{Creep tests}
\label{sec:creep}

\subsubsection{Analysis of a typical creep test}
\label{sec:singlecreep}

\begin{figure*}
\includegraphics[width=0.9\textwidth]{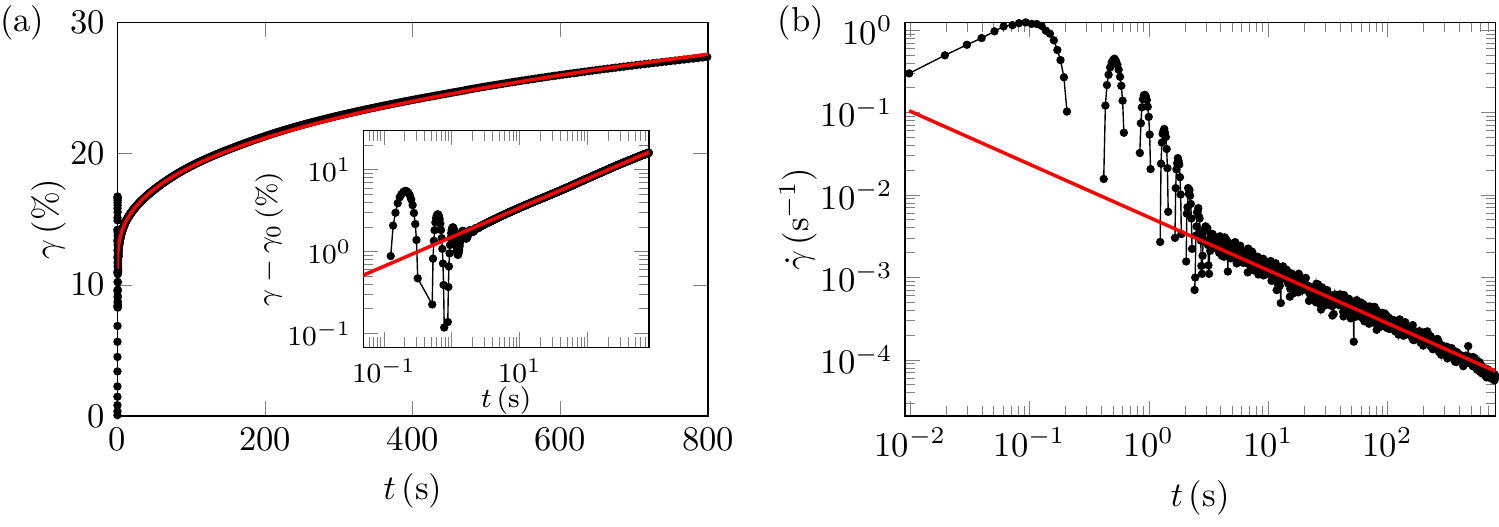}
\caption{(a)~Strain response $\gamma(t)$ for a creep experiment at $\sigma=\SI{5}{\pascal}$ (preshear at $\dot{\gamma}_\text{p}=\SI{100}{\per\second}$ for $t_\text{p}=\SI{60}{\second}$, rest time $t_\text{w}=\SI{300}{\second}$). The red solid line is the best Andrade fit, $\gamma(t)=\gamma_0 + (t/\tau)^\alpha$, with $\gamma_0 = 11.2~\%$, $\alpha=0.36$ and $\tau=\SI{1.3e5}{\second}$, following the procedure described in the text. The inset shows $\gamma - \gamma_0$ with $\gamma_0 = 11.2~\%$ as a function of time in logarithmic scales. (b)~Shear rate response $\dot{\gamma}(t)$ corresponding to the experiment in (a). The red solid line shows the best power-law fit with exponent $\alpha-1=-0.64$.}
\label{fig:creep_fit}
\end{figure*}

Figure~\ref{fig:creep_fit}(a) displays a typical strain response $\gamma(t)$ recorded during a creep test under a shear stress $\sigma$ well below the yield stress $\sigma_\text{y}\simeq 10$~Pa but larger than 2~Pa. The short-time response involves damped oscillations [see inset of Fig.~\ref{fig:creep_fit}(a) for $t<1$~s] that are characteristic of the coupling between the instrument inertia and the material elasticity in stress-controlled rheometry \cite{Ferry:1980}. Although analyzing such oscillations may provide valuable information \cite{Baravian:1998,Baravian:2007,Benmouffok:2010,Ewoldt:2007}, we shall disregard them in the present work as we focus on the long-term creep behaviour of our microgels. In order to quantify the response to creep, we first consider the shear rate  $\dot\gamma(t)$ obtained by differentiating $\gamma(t)$ with respect to time and plotted in Fig.~\ref{fig:creep_fit}(b). As seen by using logarithmic scales, $\dot\gamma(t)$ decays as a power law of time, $\dot\gamma(t)\sim t^{\alpha-1}$ with $\alpha-1\simeq -0.64$ over almost three decades in time.

Based on the value $\alpha$ of the exponent extracted from $\dot\gamma(t)$, we then go back to the strain data and fit $\gamma(t)$ by a power law: 
\begin{equation}
\gamma(t)=\gamma_0 + (t/\tau)^\alpha\,,
\label{eq:strain}
\end{equation}
with only two adjustable parameters, namely the initial strain $\gamma_0$ and the prefactor $\tau$. Such a fit is restricted to times larger than a few seconds in order to exclude short-time oscillations. The result is shown as a red line in Fig.~\ref{fig:creep_fit}(a) and the inset confirms that $\gamma(t)-\gamma_0$ grows as a power law of time. Such a power law is strikingly similar to the power-law creep with exponent 1/3 first reported by Andrade in metallic wires stretched under a constant load \cite{Andrade:1910}. This behaviour, referred to as ``Andrade-like'' scaling, will be discussed in more details in Sect.~\ref{sec:andrade} below.

The initial strain $\gamma_0$ corresponds to the instantaneous, elastic deformation that would be observed in the absence of inertio-elastic oscillations i.e. in an ``ideal'' creep experiment. The time $\tau$ in Eq.~(\ref{eq:strain}) corresponds to the point where the strain has increased by 100~\% starting from $\gamma_0$. However, $\tau$ cannot be interpreted as some ``characteristic time'' of the material since it appears in a power law. As a matter of fact, $\tau$ is generally much larger than the creep duration itself ($\tau=\SI{1.3e5}{\second}$ in the present case) and has no specific physical significance. In the following, we shall thus discuss only the exponent $\alpha$ and the strain $\gamma_0$. 


\subsubsection{Creep dependence on the applied stress}
\label{sec:multiplecreep}

\begin{figure}
\includegraphics[width=0.45\textwidth]{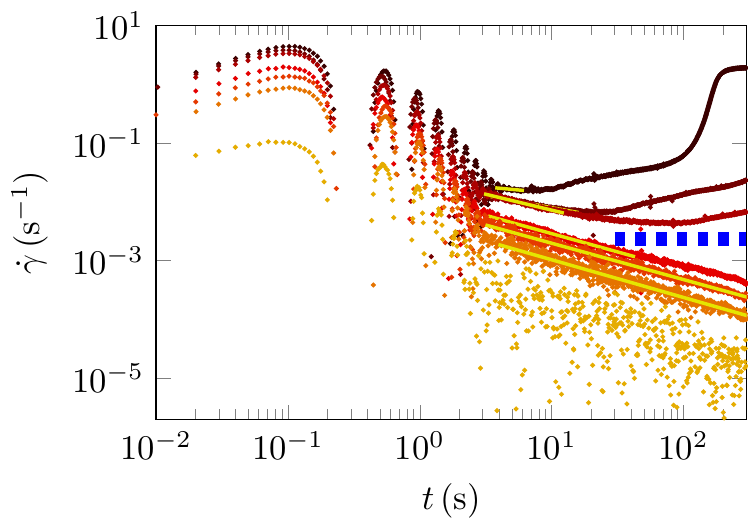}
\caption{Shear rate responses $\dot{\gamma}(t)$ for creep experiments at $\sigma=\SI{15}{}$, \SI{14}{}, \SI{12}{}, \SI{7.4}{}, \SI{5.0}{}, \SI{3.4}{} and \SI{1.0}{\pascal} from top to bottom ($\dot{\gamma}_\text{p}=\SI{100}{\per\second}$, $t_\text{p}=\SI{60}{\second}$, $t_\text{w}=\SI{300}{\second}$). The yellow solid lines show the best power-law fits of $\dot{\gamma}(t)$ over the corresponding fitting intervals (see text). The negative parts of the short-time oscillations do not show due to the use of logarithmic scales. Blue dashes separate creep experiments at $\sigma>\sigma_\text{y}$ from those at $\sigma<\sigma_\text{y}$.}
\label{fig:creep_regimes}
\end{figure}

Shear rate responses to several creep tests for imposed stresses ranging from 1~Pa to 15~Pa are gathered in Fig.~\ref{fig:creep_regimes}. While experiments performed above the yield stress $\sigma_\text{y}\simeq 10$~Pa show a transition from a creep regime to a flowing regime with a non-zero steady-state shear rate (see dark brown curve for $\sigma=15$~Pa) or at least an upturn in the shear rate (see curves for $\sigma=12$ and 14~Pa), creep tests for $\sigma<\sigma_\text{y}$ all decrease to vanishingly small shear rates. This behaviour is typical of the yielding transition. Below the yield stress, the system slowly creeps and the shear rate tends towards zero. Above the yield stress, the material eventually flows although such a solid-to-fluid transition takes increasingly long times as the applied stress approaches the yield stress.

The creep behaviour of microgels made of carbopol ETD~2050  --yet with a different preparation protocol-- has been extensively studied {\it  above} the yield stress by Divoux {\it et al.} \cite{Divoux:2011a,Divoux:2012}. It was shown that the fluidisation time diverges as a power law of the viscous stress and that the transition towards a flowing state involves a solid--fluid coexistence through the presence of long-lived transient shear bands. Although the goal of the present study is to focus on creep {\it below} the yield stress, we emphasize that an Andrade-like scaling was already reported in \cite{Divoux:2011a} with an exponent of about 1/3 and that the creep regime was shown to involve a locally homogeneous deformation field at scales larger than a few tens of micrometers thanks to ultrasonic velocimetry.

\begin{figure*}
\includegraphics[width=0.9\textwidth]{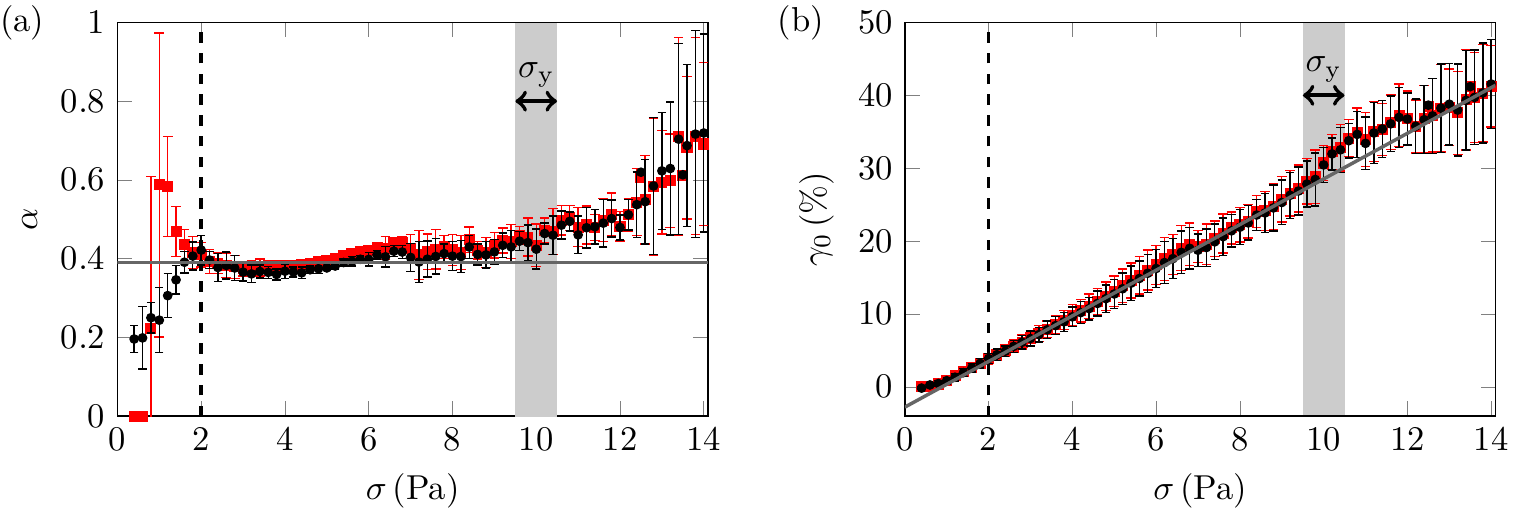}
\caption{Andrade fit parameters, (a)~exponent $\alpha$ and (b)~initial elastic deformation $\gamma_0$, as a function of the applied stress $\sigma$. The data are averaged over three to six independent creep experiments on the same carbopol batch but for different loadings. Error bars correspond to the standard deviation of the Andrade parameter estimations. Black dots correspond to estimations based on power-law fits of the shear rate response (yielding $\alpha$) then on fits of the strain response with two free parameters (yielding $\gamma_0$ and $\tau$) while red squares correspond to full fits of the strain response given by Eq.~(\ref{eq:strain}) with three free parameters. The gray line in (a) shows the mean value of $\langle\alpha\rangle=0.39$ (average over $\sigma=1.8$--8.6~Pa). The gray line in (b) is a linear fit $\sigma=G'_0\gamma_0+\sigma_0$ with $G'_0=\SI{32}{\pascal}$ and $\sigma_0=\SI{0.9}{\pascal}$ (fit over $\sigma=2$--10~Pa). The gray area shows the range of yield stresses $\sigma_\text{y}$ found by systematically fitting the flow curve (see Fig.~\ref{fig:flowcurve}) for the various loadings prior to creep experiments. The vertical dashed line shows the typical stress of 2~Pa below which residual stresses significantly affect the creep response (see Sect.~\ref{sec:interplay}).}
\label{fig:andrade}
\end{figure*}

The parameters $\alpha$ and $\gamma_0$, extracted from power-law fits of $\dot{\gamma}(t)$ and $\gamma(t)$ as described above, are plotted as a function of the imposed stress $\sigma$ with black symbols in Fig.~\ref{fig:andrade}. The time intervals over which the fits are performed depend on $\sigma$ and are indicated by yellow lines in Fig.~\ref{fig:creep_regimes}. The exponent $\alpha$ is seen to remain roughly constant for $0.2 \sigma_\text{y}\lesssim\sigma\lesssim \sigma_\text{y}$ and its mean is $\langle\alpha\rangle=0.39$. For stresses above the yield stress, the exponent seems to increase significantly. This could be due to the fact that for $\sigma\gtrsim \sigma_\text{y}$, the amplitude of the initial inertio-elastic oscillations may reach into the nonlinear regime and significantly affect the material, e.g. through microstructural rearrangements. Yet, the fitting interval gets smaller as the creep regime becomes shorter under larger applied stresses, which makes the error bars on $\alpha$ increase dramatically. The value of the Andrade exponent $\alpha$ for $\sigma\lesssim \sigma_\text{y}$ will be further discussed in Sect.~\ref{sec:andrade} in light of the current literature. 

Not unexpectedly, the initial elastic deformation $\gamma_0$ increases linearly with $\sigma$ over the whole range of applied shear stresses and the proportionality factor $G'_0=32$~Pa is fully consistent with the elastic modulus $G_0=34$~Pa measured independently through small oscillatory shear. However, it is important to note that the linear fit shown in Fig.~\ref{fig:andrade}(b) involves a non-negligible value for $\sigma=0$. In other words, one has $\sigma=G'_0\gamma_0+\sigma_0$ where $\sigma_0=0.9$~Pa is significant and points to a specific behaviour at low applied stresses. Correspondingly, the exponent $\alpha$ sharply deviates from its mean value when the stress is decreased below a characteristic value $\sigma_\text{c}\simeq 2$~Pa.

\begin{figure}
\includegraphics[width=0.45\textwidth]{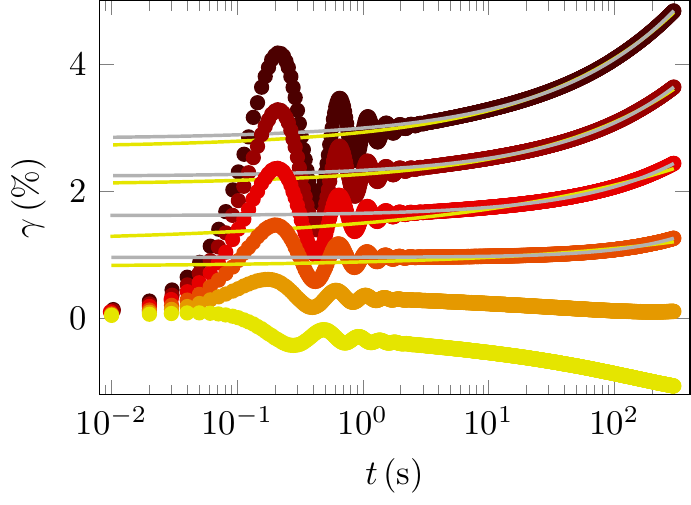}
\caption{Strain responses $\gamma(t)$ for creep experiments at low stresses, $\sigma=\SI{0.6}{},\SI{0.8}{},\SI{1.0}{},\SI{1.2}{},\SI{1.4}{},\SI{1.6}{\pascal}$, from bottom to top ($\dot{\gamma}_\text{p}=\SI{100}{\per\second}$, $t_\text{p}=\SI{60}{\second}$, $t_\text{w}=\SI{300}{\second}$). The yellow lines are the power-law fits obtained with the procedure described in Sect.~\ref{sec:singlecreep} and leading to exponents $\alpha=0.29$, 0.23, 0.38 and 0.40 from bottom to top. The gray lines correspond to the best fits of $\gamma(t)$ by Eq.~\ref{eq:strain} with three adjustable parameters and yield $\alpha=0.84$, 0.54, 0.47 and 0.44 from bottom to top.}
\label{fig:low_stress1}
\end{figure}

To test the robustness of the above estimates, Fig.~\ref{fig:andrade} also displays the parameters $\alpha$ and $\gamma_0$ obtained by fitting directly the strain response to Eq.~(\ref{eq:strain}) with three free parameters. Over the range $\sigma=2$--10~Pa, the results are almost undistinguishable from those obtained with the previous fitting procedure. For $\sigma\lesssim \sigma_\text{c}$ however, the two estimates of $\alpha$ strongly differ. This may be due to the fact that for low stresses, differentiating $\gamma(t)$ leads to noisy shear rate responses involving negative values of $\dot\gamma(t)$ which are not taken into account when fitting the logarithm of the data to a straight line (see Fig.~\ref{fig:creep_regimes} for $\sigma=1$~Pa). This also suggests that Eq.~(\ref{eq:strain}) may fail in describing the strain response for $\sigma\lesssim\sigma_\text{c}$.

In order to get better insight into the creep behaviour at low applied stresses, Fig.~\ref{fig:low_stress1} displays strain responses recorded for $\sigma<2$~Pa in semilogarithmic scales. For $1\le\sigma<2$~Pa, $\gamma(t)$ can still be well fitted to power laws, leading to the values of $\alpha$ reported in Fig.~\ref{fig:andrade}(a). It could be tempting to interpret the low values of $\alpha$ obtained by first fitting $\dot\gamma(t)$ (yellow solid lines in Fig.~\ref{fig:low_stress1}) as logarithmic creep. However, fitting $\gamma(t)$ with three free parameters over the same time range leads to much larger values of $\alpha$ (gray solid lines in Fig.~\ref{fig:low_stress1}). This is indicative of a high sensitivity to the fitting procedure and questions the validity of the power-law model. Even more strikingly, at even lower stresses ($\sigma<1$~Pa), strain responses no longer show a monotonic increase once the initial oscillations have died out and $\gamma(t)$ even goes to negative values for the smallest applied stress of 0.6~Pa. As shall be shown in the next section, this anomalous behaviour is the signature of a subtle interplay between creep and residual stresses following preshear.

\subsection{Interplay between creep and residual stresses}
\label{sec:interplay}

\subsubsection{Evidence for residual stresses}
\label{sec:residual}

\begin{figure*}
\includegraphics[width=0.9\textwidth]{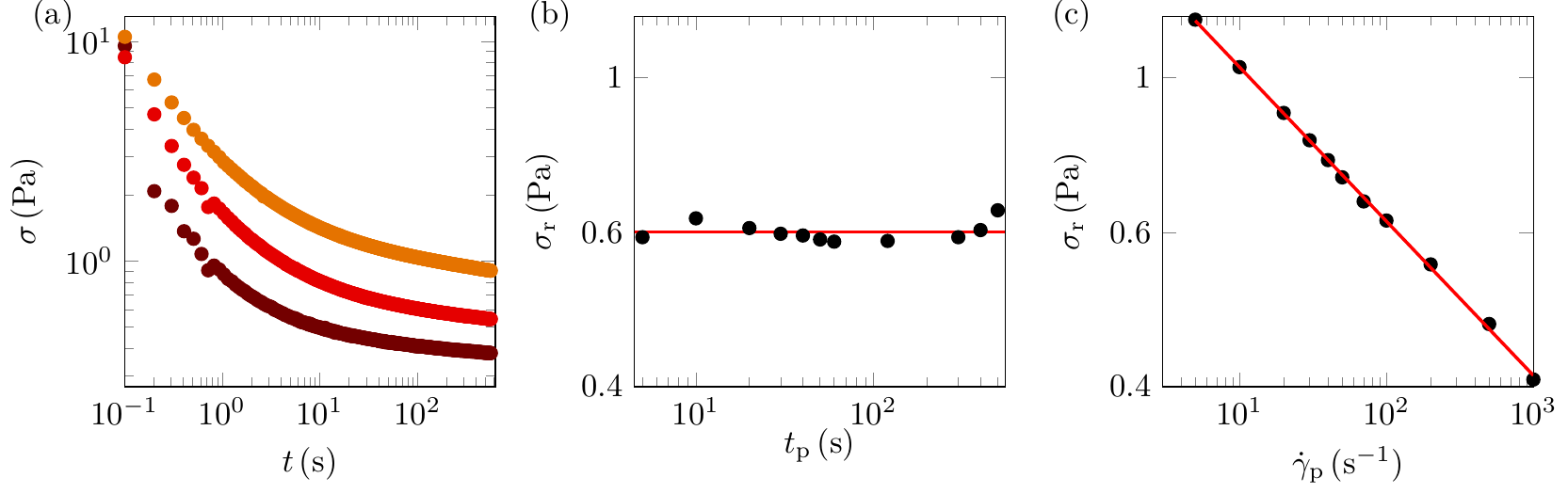}
\caption{(a)~Stress relaxation $\sigma(t)$ after preshear at $\dot{\gamma}_\text{p}=$10, 100 and \SI{1000}{\per\second} (from top to bottom) for $t_\text{p}=\SI{60}{\second}$. A zero shear rate is imposed immediately after preshear. (b)~Residual stress $\sigma_\text{r}$, defined as the stress after a relaxation over $t_\text{w}=\SI{600}{\second}$, as a function of the preshear duration $t_\text{p}$ for a fixed $\dot{\gamma}_\text{p}=\SI{100}{\per\second}$. The red line is $\langle \sigma_\text{r} \rangle = \SI{0.63}{\pascal}$.  (c)~$\sigma_\text{r}$ as a function of the preshear rate $\dot{\gamma}_\text{p}$ for a fixed $t_\text{p}=\SI{60}{\second}$. The red line is the best power-law fit $\sigma_\text{r} = 1.6 \dot{\gamma}_\text{p}^{-0.2}$.}
\label{fig:residual}
\end{figure*}

In order to interpret the unusual creep behaviour under low applied stresses, we now turn to the flow cessation experiments shown in Fig.~\ref{fig:residual}(a). The stress $\sigma(t)$ is seen to slowly relax and does not seem to tend to zero at long times. This indicates that internal stresses remain within the material over long time scales. For obvious practical reasons, we have to stop the relaxation after a given waiting time $t_\text{w}$, here $\SI{600}{\second}$. We then define the ``residual stress''  simply as $\sigma_\text{r}=\sigma(t_\text{w})$. This residual stress is of the order of 1~Pa and decreases with the preshear rate $\dot{\gamma}_\text{p}$. Figures~\ref{fig:residual}(b) and (c) show that for a given $\dot{\gamma}_\text{p}$, $\sigma_\text{r}$ is independent of the preshear duration $t_\text{p}$ but that it decreases as $\dot{\gamma}_\text{p}^{-0.2}$. This means that the material only keeps a memory of the preshear intensity (as long as it has been previously thoroughly fluidised by preshear).

As will be discussed in more details below in Sect.~\ref{sec:discuss_residual}, residual stresses result from the quench from a high shear rate to zero shear rate, which traps the microstructure into a configuration that slowly relaxes over time. Here, we note that the values taken by $\sigma_\text{r}$ roughly correspond to the stress $\sigma_\text{c}$ below which the Andrade exponent significantly differs from $\langle\alpha\rangle=0.39$ in Figs.~\ref{fig:andrade}(a) and \ref{fig:low_stress1}. It is also of the same order of magnitude as the value $\sigma_0=0.9$~Pa found in Fig.~\ref{fig:andrade}(b). Therefore, the anomalous creep behaviour observed at low stress most likely arises from an effect of this residual stress. 

\subsubsection{Interpretation of anomalous creep at low applied stresses}
\label{sec:creep_low}

\begin{figure*}
\includegraphics[width=0.9\textwidth]{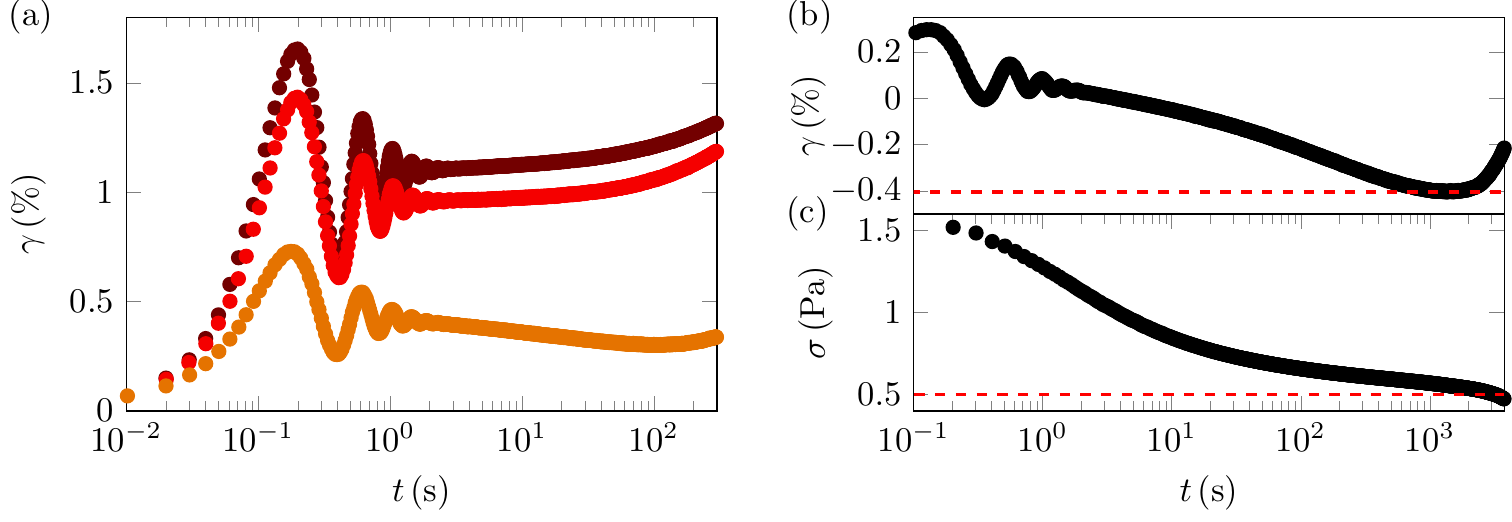}
\caption{(a)~Strain responses $\gamma(t)$ for creep experiments at $\sigma=\SI{1}{\pascal}$ after preshear at $\dot{\gamma}_\text{p}=$5, 20, $\SI{1000}{\per\second}$ from bottom to top ($t_\text{p}=\SI{60}{\second}$, $t_\text{w}=\SI{300}{\second}$). (b)~Strain response $\gamma(t)$ over one hour for a creep experiment at $\sigma=\SI{0.5}{\pascal}$ ($\dot{\gamma}_\text{p}=\SI{100}{\per\second}$, $t_\text{p}=\SI{60}{\second}$, $t_\text{w}=\SI{300}{\second}$).
(c)~Stress relaxation $\sigma(t)$ immediately after preshear ($\dot{\gamma}_\text{p}=\SI{100}{\per\second}$, $t_\text{p}=\SI{60}{\second}$). Same loading as in (b). Note that the origin of time corresponds to 300~s after preshear in (a,b) while preshear ends at $t=0$ in (c).}
\label{fig:low_stress2}
\end{figure*}

To confirm the link between anomalous creep and residual stresses, we go back to creep experiments at low imposed stress but we now vary the preshear rate $\dot{\gamma}_\text{p}$ for a fixed $\sigma=1$~Pa in Fig.~\ref{fig:low_stress2}(a). The strain response is observed to decrease initially only for the lowest preshear rate i.e. for the largest residual stress after the rest period of 300~s following preshear. When the preshear rate is increased such that the corresponding residual stress lies significantly below the applied stress, Andrade-like creep is recovered.

Figures~\ref{fig:low_stress2}(b) and (c) focus on two one-hour long experiments performed on the same loading, respectively a creep test at a very low stress of 0.5~Pa and a stress relaxation for the same preshear protocol. Interestingly, the strain $\gamma(t)$ in Fig.~\ref{fig:low_stress2}(b) decreases as long as the stress $\sigma(t)$ is larger than 0.5~Pa in Fig.~\ref{fig:low_stress2}(c) and the time at which $\sigma(t)$ reaches 0.5~Pa is of the same order of magnitude as the time at which $\gamma(t)$ goes through a minimum value of about $-0.4\%$ and starts increasing (see red dashes).

Our interpretation of anomalous creep at low stress is thus as follows. When brought to rest at a zero shear rate, the presheared material stores some internal stresses that do not fully relax even on very long time scales. After a waiting time of 300~s, the resulting macroscopic stress $\sigma_\text{r}$ is typically about 1~Pa. When a stress $\sigma$ is subsequently applied, the material actually ``feels'' an effective stress $\sigma-\sigma_\text{r}$. When $\sigma<\sigma_\text{r}$, this results in a decreasing strain response after the initial elastic deformation. During such a ``backward creep,'' the strain may then reach negative values. To put it differently, relaxing residual stresses tend to deform the material in the direction opposite to that of preshear. This is most probably because microstructural deformations induced by preshear in the forward direction are quenched during the rest time at zero shear rate so that releasing the stress (or applying $\sigma<\sigma_\text{r}$) tends to relax the structure in the reverse direction. As the microstructure slowly relaxes, the applied stress eventually overcomes residual stresses and the strain $\gamma(t)$ starts increasing, although this may take very long times. On the other hand, when $\sigma\gtrsim\sigma_\text{r}$, the strain increases right from the start of the creep experiment but residual stresses slow down the strain response, leading to the flatter $\gamma(t)$ observed for $1<\sigma<2$~Pa. This competition between structural relaxation and the applied stress leads to the anomalous creep response observed for $\sigma<\sigma_\text{c}\simeq 2$~Pa. Long-time behaviours similar to those reported here in Figs.~\ref{fig:low_stress1} and \ref{fig:low_stress2} have been observed previously in other microgels under low stresses \cite{Cloitre:2000,Purnomo:2007} and raise the question of aging. This will be further discussed in Sect.~\ref{sec:aging}.

\section{Discussion}
\label{sec:discussion}

Our main findings concern (i)~the evidence for robust power laws that characterize the creep behaviour of ETD~2050 carbopol microgels below the yield stress $\sigma_\text{y}$ and down to $\sigma_\text{c}\simeq 0.2 \sigma_\text{y}$ and (ii)~the presence of residual stresses after a rest time under a zero shear rate following preshear. These residual stresses are of the same order of magnitude as $\sigma_\text{c}$ and compete with the applied stress in creep experiments for $\sigma\lesssim\sigma_\text{c}$ and that raise the question of long-term relaxation and glassy dynamics. As shown in Appendix~B, these findings are robust to a change of carbopol concentration. Moreover, in Appendix~C, we test the more classical protocol where a zero shear stress (rather than a zero shear rate) is imposed during the rest time. These additional measurements show that power-law creep is still present and characterized by the same exponent but that anomalous creep is no longer observed at low applied stresses. Since this zero-stress protocol cancels out any residual stress, this confirms the interplay between creep and residual stresses when a quench to a zero shear rate is applied following preshear. In this section, we further discuss our results in light of previous works on soft materials.

\subsection{Andrade-like creep in soft materials}
\label{sec:andrade}

Power-law creep has been reported in a number of soft systems including cellulose solutions \cite{Plazek:1960}, aging polymers \cite{Cheriere:1997}, hexagonal columnar phases \cite{Bauer:2006}, collagen solutions \cite{Gobeaux:2010}, carbopol microgels similar to those studied here \cite{Moller:2009a,Divoux:2011a,Dimitriou:2013}, emulsions \cite{Paredes:2013,Dinkgreve:2015}, colloidal glasses \cite{Siebenburger:2012,Chan:2014}, colloidal gels made of attractive carbon black particles \cite{Grenard:2014}, natural gum \cite{Jaishankar:2012}, protein gels \cite{Brenner:2009,Brenner:2013,Leocmach:2014} and living cells \cite{Desprat:2005,Balland:2006,Kollmannsberger:2011,Hecht:2015}. The exponent $\alpha$ ranges from 0.2 to 0.7 depending on the system. Therefore, Andrade-like creep is ubiquitous in soft matter and the main open question is how it can be interpreted in terms of microstructure.

It appears that the various materials listed above can be broadly divided into two categories. A first category encompasses biological and/or fibrous-like materials such as living cells, protein, collagen or cellulose gels, whose creep eventually leads to irreversible failure. These materials are generally characterized not only by power-law creep but also by a ``power-law rheology'' in the sense that their linear viscoelastic moduli depend upon frequency as power laws, $G'\sim G''\sim f^\alpha$, with the same exponent as the strain in creep experiments. Power-law rheology is generally understood as deriving from the fractal nature of the underlying microstructural network \cite{Muthukumar:1989,Patricio:2015,Hung:2015}. Since the viscoelastic spectrum and the creep compliance are related through a Laplace transform, the link between power-law rheology and power-law creep is straightforward at least in the small deformation regime  \cite{Desprat:2005,Tschoegl:2012}. Interestingly, based on Scott-Blair's fractional approach and material ``quasiproperties'' \cite{Blair:1947}, such a link can be extended to more complex linear rheologies and even to nonlinear viscoelasticity \cite{Jaishankar:2014}. Moreover, power-law creep, as well as the critical behaviour of some protein gels approaching failure \cite{Leocmach:2014}, is well captured by a family of models known as ``fiber-bundle models'' (FBMs) \cite{Kun:2003b,Kovacs:2008,Jagla:2011} that were originally devised for solid fibrous materials such as fiber-matrix composites \cite{Nechad:2005a,Nechad:2005b}. However, so far, some important assumptions of FBMs, such as local \textit{vs} global load sharing \cite{Kun:2003a} or damage accumulation \cite{Halasz:2012} by the ``fibers'' that constitute the microstructure, have not been evidenced experimentally in real soft systems.

The second category, which is more relevant to the present work, gathers a number of ``soft glassy materials'' that are characterized by an amorphous microstructure made of a dense assembly of particles. These materials include hexagonal columnar phases, colloidal glasses, emulsions and carbopol microgels. They generically show an almost frequency-independent elastic modulus so that there is not such an obvious link between their linear viscoelasticity and power-law creep as for the previous class of materials. Rather, the origin of power-law creep in soft glassy materials and its relation to their structural properties remain to be fully unveiled. In particular, a recent single-mode reduction of mode-coupling theory (MCT) generalized to transient regimes suggests that under certain conditions and in the low-frequency range, the elastic modulus is essentially frequency-independent and the viscous modulus scales as $G''\sim f^\alpha$ where $\alpha$ corresponds to the Andrade exponent \cite{Frahsa:2015}. Here, we note from Fig.~\ref{fig:oscill}(a) that $G''\sim f^{0.44}$ so that this exponent is indeed compatible with the Andrade exponent. Clearly, more systematic work is required to decide whether this is mere coincidence or whether there is a fundamental connection between the frequency-dependence of the viscous modulus and power-law creep in soft glassy materials. Let us also point out that a recent model of an elasto-plastic material with kinematic hardening predicts power-law creep where the Andrade exponent $\alpha$ is linked to the Herschel-Bulkley exponent $n$ through $\alpha=n/(n-1)$ for $\sigma<\sigma_\text{y}$ and $\alpha=0$ for $\sigma=\sigma_\text{y}$ \cite{Dimitriou:2013}. Unfortunately, with $0<n<1$, this corresponds to negative values of $\alpha$ i.e. to decreasing strains inconsistent with the Andrade-like creep observed experimentally.

By analogy with the motion of dislocations and vacancies that gives rise to Andrade creep in crystalline solids \cite{Cottrell:1952,Mott:1953,Cottrell:1997,Miguel:2002,Miguel:2008,Nabarro:2004}, it has been speculated that power-law creep arises from collective particle dynamics within the dense, disordered microstructure. A phenomenological model based on this idea and introducing a time-dependent fraction of mobile and arrested particles correctly predicts the Andrade exponent $\alpha\simeq 0.4$ observed in emulsions from the scaling of the steady-state flow curve with the volume fraction \cite{Paredes:2013}. Direct evidence for such heterogeneous dynamics under creep came only very recently from both simulations and experiments on hard-sphere colloidal glasses \cite{Chaudhuri:2013,Sentjabrskaja:2015}. Using simultaneous rheology and confocal microscopy, Sentjabrskaja {\it et al.} could make a quantitative link between the strain response $\gamma(t)$ and dynamical heterogeneities that remain spatially localized and sub-diffusive during creep but grow in size and become transiently super-diffusive at the onset of steady flow. In particular, they showed that the mean squared displacement $\Delta^2(t)$ follows the same power law as the strain, $\Delta^2(t)\sim\gamma(t)\sim t^\alpha$, with $\alpha\simeq 0.4$ for $\sigma=0.9\sigma_\text{y}$ in simulations and $\alpha\simeq 0.5$ for $\sigma\simeq\sigma_\text{y}$ in experiments. Although applied stresses smaller than the yield stress were not investigated experimentally, probably due to the very small value of  $\sigma_\text{y}\simeq 10$~mPa, this study provides promising key results on microscopic dynamics during the creep of colloidal glasses. Interestingly, simulations show that dynamical heterogeneities may arrange into macroscopic shear bands during power-law creep \cite{Chaudhuri:2013} while local measurements --yet on rather large spatial scales-- have suggested that the deformation remains homogeneous \cite{Divoux:2011a,Grenard:2014}. This obviously deserves more work and confocal microscopy appears as an excellent tool to address this issue.

Of course the present measurements, which are restricted to global rheological data, do not provide any insight into local dynamics. However, we note that the Andrade exponent takes very similar values in carbopol, emulsions and colloidal glasses. Our results also show that power-law creep is observed below the yield stress down to very small applied stresses. We believe that microscopy experiments coupled to rheology are in line to fully understand creep in carbopol microgels where the above picture of mobile \textit{vs} arrested particles is likely to fail since the microstructure is constituted of swollen polymer particles that are compressed and deformed against each other.
 
\subsection{Residual stresses in soft materials}
\label{sec:discuss_residual}

Over the last few years, interest into residual stresses in soft materials has grown spectacularly. It is acknowledged that residual stresses (also sometimes referred to as ``internal'' stresses) result from history-dependent microstructural features that are frozen-in upon cessation of shear and generally decrease with the preshear rate or stress. Recent studies include polyelectrolyte microgels \cite{Mohan:2013,Mohan:2015}, colloidal gels \cite{Osuji:2008b,Negi:2009b,Negi:2010c} as well as colloidal glasses \cite{Ballauff:2013}. Numerical simulations of random jammed assemblies of non-Brownian elastic spheres have shown that residual stresses can be attributed to quenched angular distortions of the microstructure that slowly relax over time \cite{Mohan:2013}.

\begin{figure*}
\includegraphics[width=0.9\textwidth]{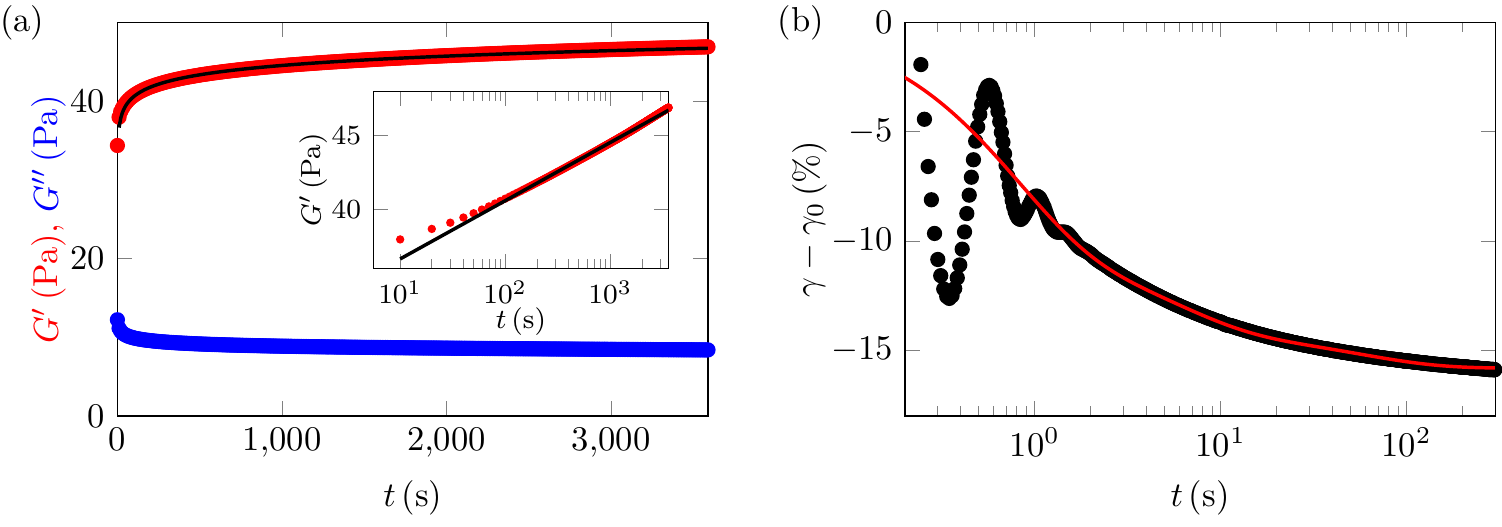}
\caption{(a)~Time evolution of the elastic modulus $G'$ (red upper curve) and of the viscous modulus $G''$ (blue bottom curve) of a 1~\%~wt carbopol microgel after preshear at $\dot{\gamma}_\text{p}=\SI{100}{\per\second}$ for $t_\text{p}=\SI{60}{\second}$. $G'$ and $G''$ are recorded under a small-amplitude oscillatory shear of strain amplitude $\gamma=1~\%$ and frequency $f=\SI{1}{\hertz}$. The black solid line is the best logarithmic fit of the elastic modulus, $G'(t)=G_1+G_2\log(t)$, with $G_1=\SI{32.7}{\pascal}$ and $G_2=\SI{3.91}{\pascal}$. Inset: $G'(t)$ in semilogarithmic scales. (b)~Strain relaxation $\gamma(t)$ following flow cessation in a 1~\%~wt carbopol microgel. A preshear stress $\sigma=\SI{20}{\pascal}$, much larger than the yield stress and corresponding to a steady-state shear rate of about $\SI{5}{\per\second}$, is applied for $\SI{300}{\second}$. The system is then quenched to $\sigma=0$ at $t=0$. The red solid line is the best fit by a sum of three exponential functions with relaxation times 0.7, 5.3 and $\SI{56}{\second}$. The initial value $\gamma_0$ is inferred from this fit.}
\label{fig:aging}
\end{figure*}

To the best of our knowledge, in carbopol microgels, residual stresses have only been mentioned by a few studies and in the specific contexts of bubble rise \cite{Piau:2007,Mougin:2012}, penetrometry \cite{Boujlel:2012a}, capillary rise \cite{Geraud:2014} and surface tension measurements \cite{Jorgensen:2015}. While Coussot {\it et al.} have raised the possibility that residual stresses interfere with the creep response at low applied stresses in bentonite, mustard and a hair gel mostly constituted of carbopol \cite{Coussot:2006}, no quantitative analysis of this phenomenon had been performed as in the present work. As seen in Fig.~\ref{fig:residual}(c) for a fixed $t_\text{p}$, the residual stress decreases as a weak power law of $\dot{\gamma}_\text{p}$ with an exponent $-0.2$ which seems rather insensitive to the concentration [see Fig.~\ref{fig:concentration}(c) in Appendix~B]. Interestingly, residual stresses measured from experiments on colloidal glasses and from molecular dynamics simulations are consistent with such a scaling, although power-law fits were not performed by the authors (see Fig.~2 in \cite{Ballauff:2013}). Moreover, it could be argued that the logarithmic scaling evidenced in polyelectrolyte microgels, $\sigma_\text{r}\sim\log(\sigma_\text{p}-\sigma_\text{y})$  with $\sigma_\text{p}$ the preshear stress \cite{Mohan:2013,Mohan:2015}, is actually compatible with the present weak power law of $\dot{\gamma}_\text{p}$.

\subsection{Aging or not aging?}
\label{sec:aging}

Andrade-like creep and/or residual stresses are commonly predicted from various non-ergodic models for glassy behaviour, including the ``Soft Glassy Rheology'' (SGR) model derived from local probabilistic descriptions \cite{Fielding:2000}, the ``Shear Transformation Zone'' (STZ) theory \cite{Falk:2011,Bouchbinder:2011}, MCT \cite{Ballauff:2013,Frahsa:2015} and more phenomenological approaches \cite{Dinkgreve:2015,Joshi:2015}. There, the progressive slowing down of the shear rate observed under creep is attributed to aging, i.e. to the fact that the system explores deeper and deeper potential wells in the complex energy landscape that results from the wealth of ever-evolving configurations accessible to the glassy system. In the case of step-strain response, aging can be defined as ``the property that a significant part of the stress relaxation takes place on time scales that grow with the age $t_\text{w}$ of the system'' \cite{Fielding:2000}. 

These observations obviously question the possible implication of aging in the interplay between creep and residual stresses observed in our experiments. As a matter of fact, decreasing strain responses under small applied stresses have been reported in polyelectrolyte microgels \cite{Cloitre:2000}. Together with logarithmic strain recovery curves after flow cessation, these peculiar creep responses have been interpreted in terms of aging phenomena. More precisely, the $\gamma(t)$ curves measured under a stress $\sigma\simeq 0.05\sigma_\text{y}$ first increase and then keep decreasing logarithmically over the course of several hours without any sign of levelling off. The time at which $\gamma(t)$ starts to decrease was shown to correspond to the ``age'' of the system, defined as the waiting time $t_\text{w}$ at rest between preshear and the creep measurement, so that all strain responses under low stresses can be collapsed by plotting $\gamma(t)-\gamma(t_\text{w})$ as function of $(t-t_\text{w})/t_\text{w}$. Similar results were later reported in thermosensitive pNIPAM microgels for $\sigma\simeq0.1\sigma_\text{y}$ \cite{Purnomo:2007} and successfully modeled with an STZ theory \cite{Bouchbinder:2011}. Such a striking behaviour, characterized by a response in the direction opposite to the preshear flow, was ascribed to structural relaxation and strong aging effects: after a rest time $t_\text{w}$, configurations with relaxation times faster than $t_\text{w}$ have relaxed so that, when a small stress is applied, the strain increases at short times but decreases for $t\gtrsim t_\text{w}$ as configurations with relaxation times longer than $t_\text{w}$ relax. Above a typical stress $\sigma_\text{c}\simeq 0.2\sigma_\text{y}$, ever-increasing strain responses are observed that are still dependent on $t_\text{w}$  	due to competition between aging and partial rejuvenation. This transition between normal and anomalous creep occurs at a relative stress $\sigma_\text{c}/\sigma_\text{y}\simeq 0.2$ which is quantitatively very close to that found in our experiments. 

\begin{figure}
\includegraphics[width=0.45\textwidth]{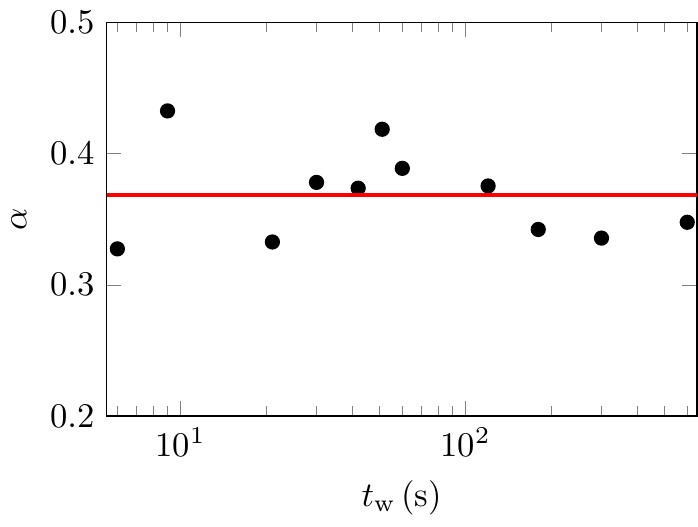}
\caption{Andrade exponent as a function of the waiting time $t_w$ after preshear for creep experiments at $\sigma=5$~Pa. The red line is $\langle \alpha\rangle = 0.37$. Experiments with $C=1$~\%~wt, $\dot{\gamma}_\text{p}=\SI{100}{\per\second}$ and $t_\text{p}=\SI{60}{\second}$.}
\label{fig:aging_andrade}
\end{figure}

In order to decide whether aging could be at play in the long-time creep behaviour of our carbopol microgels, we first test for the time evolution of the elastic modulus $G'(t)$ over one hour after preshear in Fig.~\ref{fig:aging}(a). As already reported on similar ETD~2050 carbopol samples \cite{Divoux:2011b}, we find that the elastic modulus recovers a value of about 40~Pa  within a few seconds but that it subsequently increases logarithmically with time [see inset of Fig.~\ref{fig:aging}(a)] while the viscous modulus concomitantly decreases. Therefore, we observe a slow consolidation, i.e. enhanced elasticity and smaller dissipation over time, that could be interpreted as the signature of some aging process \cite{Fielding:2000,Derec:2003,Coussot:2006}. However, focusing on strain relaxation after preshear provides a somewhat different picture. Figure~\ref{fig:aging}(b) indeed shows that following a fast quench from a stress value $\sigma=20$~Pa well above the yield stress down to zero stress, the strain $\gamma(t)$ does not decrease logarithmically as observed in polyelectrolyte microgels by \cite{Cloitre:2000}. Rather it is well fitted by a sum of three exponentials, which hints at a viscoelastic-like relaxation mechanism. Thus, it appears that our system reaches an equilibrium state within a few minutes in contrast with strongly aging materials (compare with Fig.~1 in \cite{Cloitre:2000}). We also note that in the case of thermosensitive pNIPAM microgels, strain relaxations were not strictly logarithmic and seemed to level off after a few minutes (see Fig.~4 in \cite{Purnomo:2007}). This could be indicative of aging effects intermediate between the present study and those in \cite{Cloitre:2000}. 

\begin{figure*}
\includegraphics[width=0.9\textwidth]{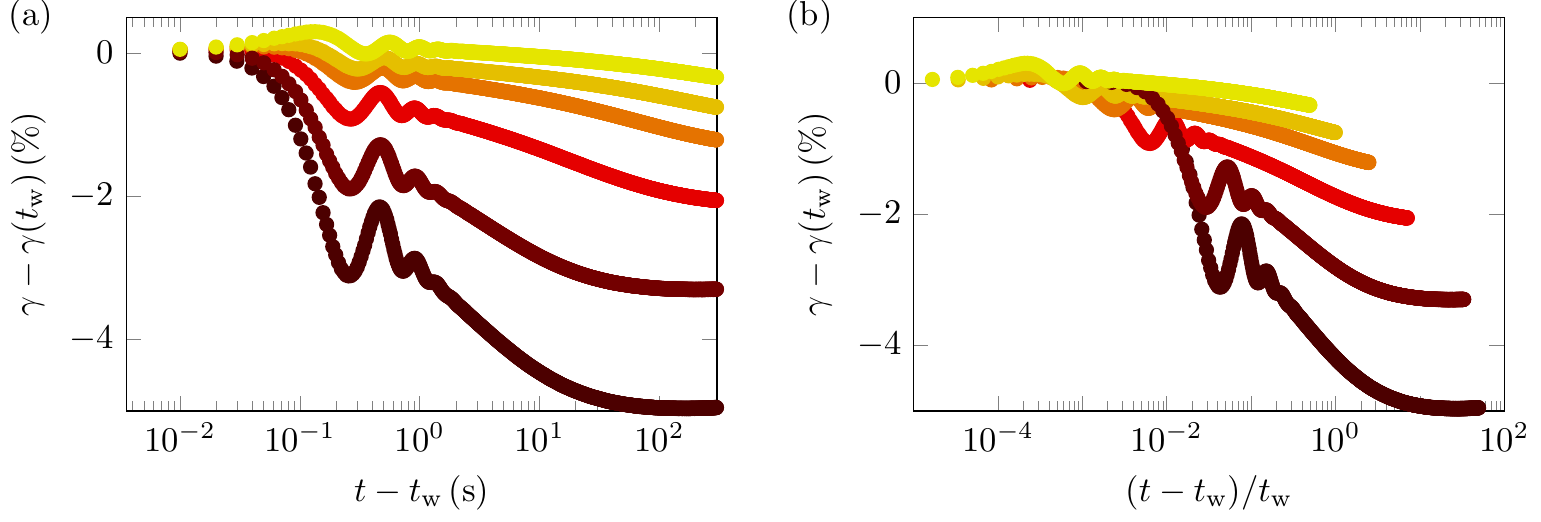}
\caption{(a)~Strain responses at $\sigma/\sigma_\text{y}=0.05$ for various wating times after preshear: $t_\text{w}=6$, 9, 42, 120, 300 and 600~s from bottom to top. Here, for the sake of consistency with previous works on aging systems \cite{Cloitre:2000,Purnomo:2007}, we explicitly denote by $t-t_\text{w}$ the time elapsed after the stress is applied at $t=t_\text{w}$. (b)~Same data plotted as a function of $(t-t_\text{w})/t_\text{w}$. Experiments with $C=1$~\%~wt, $\dot{\gamma}_\text{p}=\SI{100}{\per\second}$ and $t_\text{p}=\SI{60}{\second}$.}
\label{fig:aging_lowstress}
\end{figure*}

To further test the influence of the ``age'' of the system on its creep behaviour, we performed series of creep experiments with different rest times $t_\text{w}$ during which the viscoelastic moduli are monitored through small-amplitude oscillatory shear (strain amplitude 1~\% and frequency 1~Hz). When the applied stress is well above $\sigma_\text{c}$, we observe power-law creep whatever the age of the system. Moreover, as seen in Fig.~\ref{fig:aging_andrade} for $\sigma\simeq 0.5\sigma_\text{y}$, the Andrade exponent $\alpha$ does not display any systematic variation with $t_w$ and the average of 0.37 is consistent with $\alpha=0.39\pm 0.04$ found previously. This indicates that, if present, aging has little effect on Andrade-like creep.

In Fig.~\ref{fig:aging_lowstress}(a), a stress $\sigma=0.05\sigma_\text{y}$, well below $\sigma_\text{c}$, is applied for different values of $t_\text{w}$. There, a clear influence of $t_w$ on anomalous creep is observed: for short waiting times, the strain decrease is faster and much sharper. Such strain responses significantly differ from those of polyelectrolyte microgels (compare with Fig.~2 in \cite{Cloitre:2000}). Here, neither a strain maximum nor a long-time logarithmic decrease is observed. Moreover, as shown in Fig.~\ref{fig:aging_lowstress}(b), the strain response does not scale as $(t-t_\text{w})/t_\text{w}$. The influence of $t_\text{w}$ on $\gamma(t)$ at low stresses may thus be simply interpreted as the signature of the relaxation of residual stresses over (viscoelastic) time scales of a few tens of seconds: when  $t_\text{w}\lesssim 100$~s, Fig.~\ref{fig:residual}(a) indeed shows that residual stresses sharply decrease so that for shorter $t_\text{w}$, a higher level of residual stress induces a stronger strain decrease but the strain eventually levels off and starts to increase as the applied stress becomes larger than the residual stress.

To conclude this discussion, the above comparison with previous results on microgels suggests that aging does not have a paramount influence on the creep behaviour of our carbopol samples. Even if we observe long-lasting stress and strain relaxations, these are most likely to be of viscoelastic origin than due to true glassy dynamics. 

\section{Conclusion}
\label{sec:conclu}

We have shown that microgels constituted of carbopol ETD~2050 display power-law creep with an exponent of about 0.4 whatever the carbopol concentration. Additionally, when quenched from a fluidised state to a constant strain, these samples store residual stresses that relax in a viscoelastic-like fashion (rather than logarithmically as observed in other systems with strong aging effects). These residual stresses result in an anomalous creep behaviour under low shear stresses: for $\sigma<\sigma_\text{c}\simeq 0.2 \sigma_\text{y}$, the strain initially decreases but levels off and eventually increases after a time that depends on the preshear rate, on the rest time $t_\text{w}$ and on the applied stress. Here, we do not observe strain responses with long-time logarithmic decreasing trends that scale as $t_\text{w}$, which points to the absence of significant aging effects.

Future work will explore other types of carbopol microgels, such as carbopol Ultrez 10 and carbopol 940 or 941, in order to check for the generality of Andrade-like creep and residual stresses. Investigations based on confocal microscopy should also provide deeper insight into the origin of power-law creep and of stress relaxation in microgels.

\begin{acknowledgements}
The authors wish to thank P.~Coussot, T.~Divoux and G.~Ovar\-lez for enlightening discussions. 
\end{acknowledgements}

\section*{Appendix A: Rheological characterization of carbopol ETD~2050 samples}

\subsubsection*{Flow curve measurement}

\begin{figure}
\includegraphics[width=0.45\textwidth]{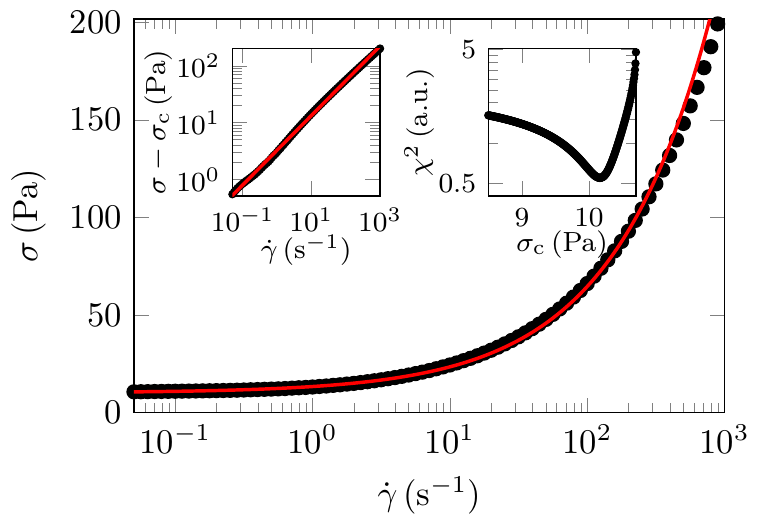}
\caption{Flow curve, shear stress $\sigma$ \textit{vs} shear rate $\dot{\gamma}$, of a 1~\% wt carbopol microgel measured in a sand-blasted cone-and-plate geometry by decreasing the shear rate in logrithmically-spaced steps of 5~s each. The red solid line is the best Herschel-Bulkley fit, $\sigma=\sigma_\text{y}+k\dot{\gamma}^n$, with $\sigma_\text{y} =10.2$~Pa, $n=0.60$ and $k=3.3$~Pa.s$^n$, inferred from the minimization procedure described in the text. The upper left inset shows $\sigma-\sigma_\text{y}$ with $\sigma_\text{y} = 10.2$~Pa as a function of $\dot{\gamma}$ in logarithmic scales. The upper right inset shows the residuals $\chi^2$ of the best power-law fit of $\sigma-\sigma_\text{y}$ \textit{vs} $\dot{\gamma}$ when varying the value of the yield stress $\sigma_\text{y}$ (see text).}
\label{fig:flowcurve}
\end{figure}

The flow curve of our 1~\% wt carbopol microgel recorded through a downward shear-rate sweep is shown in Fig.~\ref{fig:flowcurve} together with the best HB fit, $\sigma=\sigma_\text{y}+k\dot{\gamma}^n$, yielding $\sigma_\text{y} =10.2$~Pa, $n=0.60$ and $k=3.3$~Pa.s$^n$. Let us emphasize here our method for estimating the yield stress $\sigma_\text{y}$. Instead of fitting the flow curve with the full HB model that involves three free parameters, which can raise some convergence issues due to nonlinearity depending on the fitting algorithm and on the initial guess for the parameters \cite{Mullineux:2008}, we first set $\sigma_\text{y}$ to some arbitrary value and compute the best linear fit of $\ln(\sigma-\sigma_\text{y})$ \textit{vs} $\ln\dot{\gamma}$. Fitting a straight line in logarithmic coordinates rather than fitting $\sigma-\sigma_\text{y}$ \textit{vs} $\dot{\gamma}$ as a power law allows us to give the same weight to small shear rates and to larger ones. We compute the residuals $\chi^2$ of this fit, defined as the sum of the squared distances from the experimental data to the fit. We then vary $\sigma_\text{y}$ systematically over a realistic range and look for a minimum of $\chi^2$ \textit{vs} $\sigma_\text{y}$. As shown in the upper right inset of Fig.~\ref{fig:flowcurve}, $\chi^2$ goes through a well-defined minimum: the stress $\sigma_\text{y} =10.2$~Pa corresponding to this minimum is thus taken as the yield stress of our microgel. It can be checked in the upper left inset of Fig.~\ref{fig:flowcurve} that with this value of $\sigma_\text{y}$, the difference $\sigma-\sigma_\text{y}$ follows a strict power law of $\dot{\gamma}$ with no significant deviation over the whole range of shear rates. This method allows us to confidently estimate the yield stress of our microgels to within 1~\% for a given flow curve measurement. In the case of Fig.~\ref{fig:flowcurve}, this procedure yields $\sigma_\text{y} =10.2$~Pa, $n=0.60$ and $k=3.3$~Pa.s$^n$. We note that a direct HB fit of the same data with three free parameters leads to significantly different estimates ($\sigma_\text{y} =9.0$~Pa, $n=0.55$ and $k=4.5$~Pa.s$^n$), which illustrates the importance of using a careful fitting procedure as described here.

Still, it should be noted that we observe a reproducibility of the best HB fit parameters of only about 10~\%. For instance, for the same preparation batch, we found values of $\sigma_\text{y}$ ranging from 9.5 to 10.5~Pa from one loading of the cone-and-plate geometry to the other. Since the flow curves are measured through decreasing shear rate sweeps, the loading history is efficiently erased at high shear rates and such variations can only be explained by small differences in the sample volume from one loading to the other. We checked that the various flow curves are simply shifted along the stress axis from one measurement to the other and that the scatter is typically 1~Pa, which confirms that variations in HB parameters essentially stem from variations in the loaded sample volume.

\subsubsection*{Linear and nonlinear viscoelastic measurements}

\begin{figure*}
\includegraphics[width=0.65\textwidth]{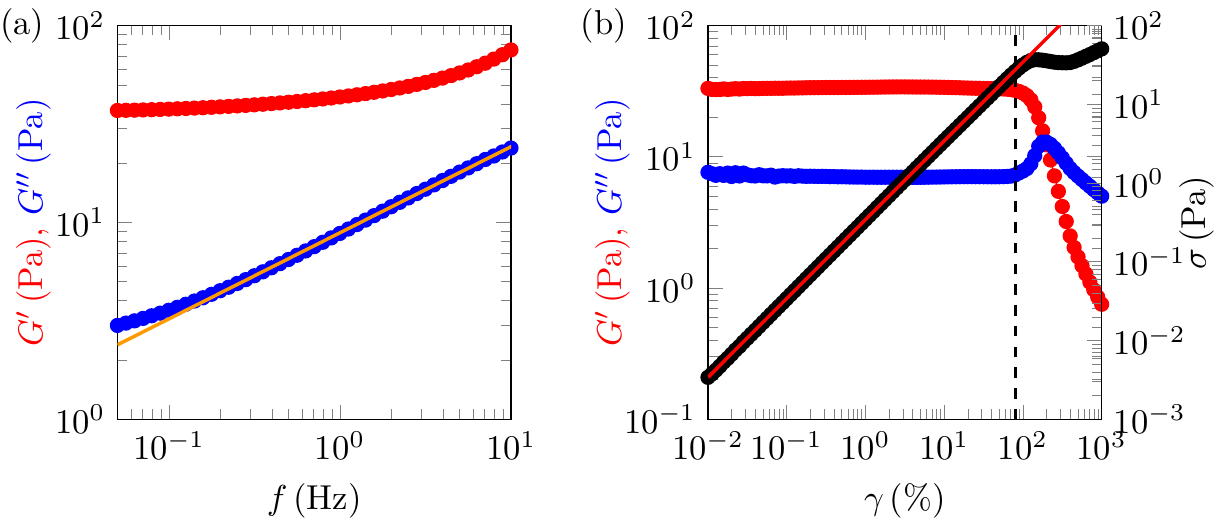}
\caption{(a)~Elastic modulus $G'$ (red) and viscous modulus $G''$ (blue) \textit{vs} frequency $f$ after a rest time $t_\text{w}=300$~s following preshear under a constant shear rate $\dot{\gamma}_\text{p}=\SI{100}{\per\second}$ for a duration $t_\text{p}=\SI{60}{\second}$. The strain amplitude is fixed to $\gamma=1~\%$ and the frequency is logarithmically swept down. The orange solid line shows that $G''\sim f^{0.44}$ for $f\gtrsim\SI{0.2}{\hertz}$. (b)~$G'$ (red, left axis), $G''$ (blue, left axis) and stress amplitude $\sigma$ (black, right axis) \textit{vs} strain amplitude $\gamma$ after the same preparation protocol as in (a). The frequency is fixed to $f=\SI{1}{\hertz}$ and the strain amplitude $\gamma$ is logarithmically swept up with a duration of 16~s per point. One has $G'(\gamma^*)=G''(\gamma^*)$ for $\gamma^*\simeq 200~\%$ which corresponds to $\sigma^*\simeq\SI{36}{\pascal}$. The red solid line is the best linear fit, $\sigma=G_0\gamma$ for $\gamma<80~\%$, leading to $G_0 = \SI{34}{\pascal}$. The dotted line indicates the end of the linear regime at $\gamma\simeq 80~\%$.}
\label{fig:oscill}
\end{figure*}

The linear viscoelastic moduli of our microgels are shown as a function of oscillation frequency $f$ in Fig.~\ref{fig:oscill}(a). The storage modulus $G'(f\rightarrow 0)=37$~Pa increases weakly over the whole frequency range. Here again, we observe variations of $G'$ by about 10~\% from one loading to the other. The loss modulus $G''$ remains always smaller than $G'$ and can be well fitted by a power law $G''\sim f^{0.44}$ at high frequencies. 
The oscillatory strain sweep of Fig.~\ref{fig:oscill}(b) shows that the linear regime, characterized by an elastic modulus $G_0=34$~Pa, extends up to strain amplitudes of about 100\%. The nonlinear regime involves a sharp drop of the storage modulus and a local maximum in the loss modulus. This corresponds to a case of ``weak strain overshoot'' as classified by Huyn {\it et al.} in their review on large-amplitude oscillatory shear \cite{Hyun:2011} and appears as a distinctive feature of soft glassy materials, including microgels \cite{deSouzaMendes:2014}. In systems like emulsions and microgels, the maximum in $G''$ is generally attributed to enhanced dissipation due to local irreversible particle rearrangements that progressively invade the whole sample before yielding and complete fluidisation \cite{Mason:1995c,Knowlton:2014}. We note that the point at which $G'$ and $G''$ cross corresponds to a stress of about 36~Pa, significantly above the yield stress measured from the flow curve in Fig.~\ref{fig:flowcurve}. Since this article is mostly devoted to creep experiments performed below the yield stress, where the strain shall not increase beyond 100~\%, we do not expand more on the nonlinear behaviour of our carbopol ETD~2050 samples and refer the reader to the cited literature for more details.

\section*{Appendix B: Influence of the carbopol concentration}

\begin{figure*}
\includegraphics[width=0.9\textwidth]{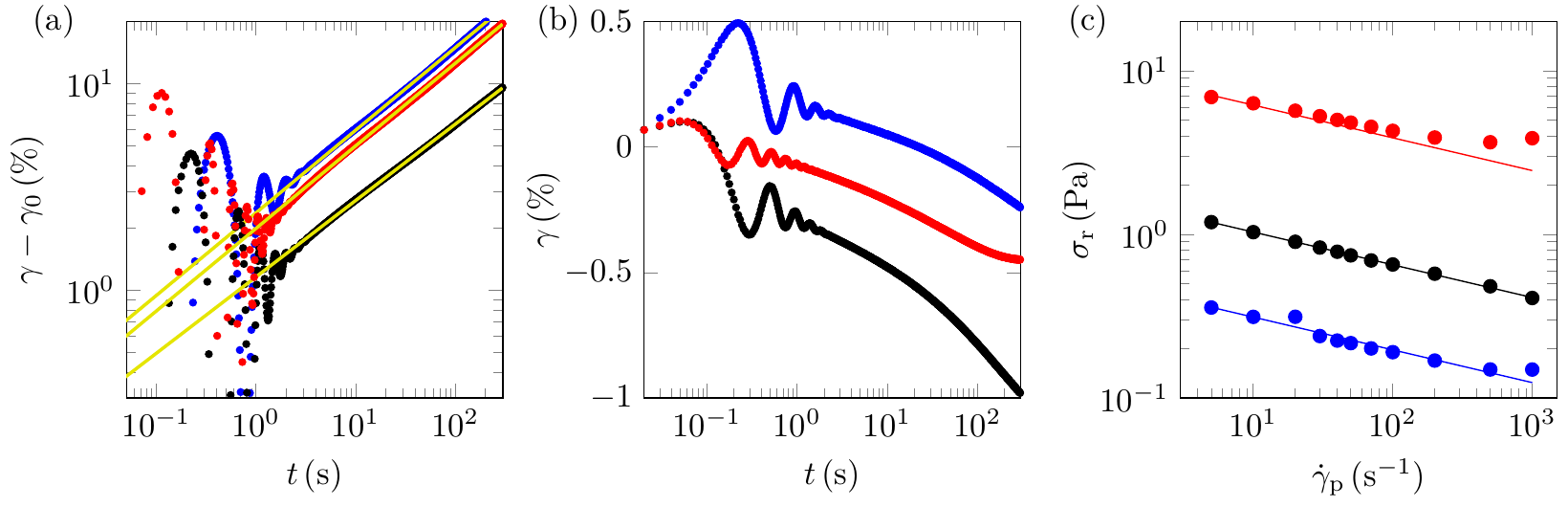}
\caption{Creep and residual stresses in carbopol microgels with concentrations $C=0.6$~\% wt (blue), 1~\% wt (black) and 2~\% wt (red). (a)~Strain responses $\gamma(t) - \gamma_0$ for  $\sigma\simeq 0.4\sigma_\text{y}$ plotted in logarithmic scales together with Andrade fits (yellow solid lines): [$C$, $\sigma/\sigma_\text{y}$, $\gamma_0$, $\alpha$] = [0.6~\% wt, 0.38, 9.9~\%, 0.40], [1~\% wt, 0.36, 8.9~\%, 0.37], [2~\% wt, 0.35, 16.2~\%, 0.40]. (b)~Strain responses $\gamma(t)$ for  $\sigma/\sigma_\text{y}\le 0.1$ plotted in semilogarithmic scales for [$C$, $\sigma/\sigma_\text{y}$] = [0.6~\% wt, 0.04], [1~\% wt, 0.04], [2~\% wt, 0.1]. (c)~Residual stress $\sigma_\text{r}$ (after a relaxation over $t_\text{w}=\SI{600}{\second}$) as a function of the preshear rate $\dot{\gamma}_\text{p}$ for $t_\text{p}=\SI{60}{\second}$. Solid lines are power laws with exponent $-0.2$.}
\label{fig:concentration}
\end{figure*}

The robustness of our findings has been tested by considering two other concentrations of carbopol ETD~2050, namely $C=0.6$~\%~wt and 2~\%~wt. As shown in Fig.~\ref{fig:concentration}(a) for similar normalized stresses, $\sigma/\sigma_\text{y}\simeq 0.4$, power-law creep is observed in all three samples. Table~\ref{tab:concentration} gathers the rheological parameters of the various samples as well as the results of the analysis of strain responses in the Andrade-like regime as described in Sect.~\ref{sec:multiplecreep}.  We find a remarkably robust mean Andrade exponent of $\alpha\simeq 0.4$ for all concentrations. Here again, the prefactor $G'_0$ deduced from the initial elastic deformation $\gamma_0$ is in good agreement with the elastic modulus $G'$. The parameter $\sigma_0$ is non-zero and increases with $C$ in the same fashion as the elastic modulus and the yield stress. This suggests that residual stresses also come into play for $C=0.6$~\%~wt and 2~\%~wt.

Indeed, for applied stresses such that $\sigma/\sigma_\text{y}< 0.1$, the same decreasing trend is observed in the strain response for all three concentrations [see Fig.~\ref{fig:concentration}(b)]. The presence of residual stresses for $C=0.6$~\%~wt and 2~\%~wt is further confirmed through flow cessation experiments performed as in Sect.~\ref{sec:residual}. Figure~\ref{fig:concentration}(c) shows that $\sigma_\text{r}$ decreases roughly as $\dot{\gamma}_\text{p}^{-0.2}$ for all concentrations with an amplitude that scales with $C$ like $G'$ and $\sigma_\text{y}$ (see Table~\ref{tab:concentration}). We conclude that a similar interplay between creep and residual stress occurs at low stresses whatever the carbopol concentration. 

\begin{table}
\caption{Elastic modulus $G'$ and yield stress $\sigma_\text{y}$ (see Appendix~A), average Andrade exponent $\langle\alpha\rangle$ and fitting parameters $G'_0$ and $\sigma_0$ (see Sect.~\ref{sec:singlecreep}), and residual stress $\sigma_\text{r}$ measured 600~s after preshear at $\dot{\gamma}_\text{p}=\SI{100}{\per\second}$ (see Sect.~\ref{sec:residual}) for carbopol microgels with different concentrations $C$.}
\label{tab:concentration}       
\begin{tabular}{clclclclclclc}
\hline\noalign{\smallskip}
$C$ (\%~wt) & $G'$ (Pa) & $\sigma_\text{y}$ (Pa) & $\langle\alpha\rangle$ & $G'_0$ (Pa) & $\sigma_0$ (Pa) & $\sigma_\text{r}$ (Pa)\\
\noalign{\smallskip}\hline\noalign{\smallskip}
0.6 & $13 \pm 1$ & $3.2 \pm 0.2$ & $0.40\pm 0.07$ & $9.8$ & $0.15$ & 0.20 \\
1 & $37 \pm 4$ & $10.0 \pm 0.5$ & $0.39\pm 0.04$ & $32$ & $0.9$ & 0.65 \\
2 & $105 \pm 10$ & $34.5 \pm 0.7$ & $0.38\pm 0.03$ & $120$ & $4.7$ & 4.9 \\
\noalign{\smallskip}\hline
\end{tabular}
\end{table}

\section*{Appendix C: Influence of the rest protocol}

\begin{figure*}
\includegraphics[width=0.9\textwidth]{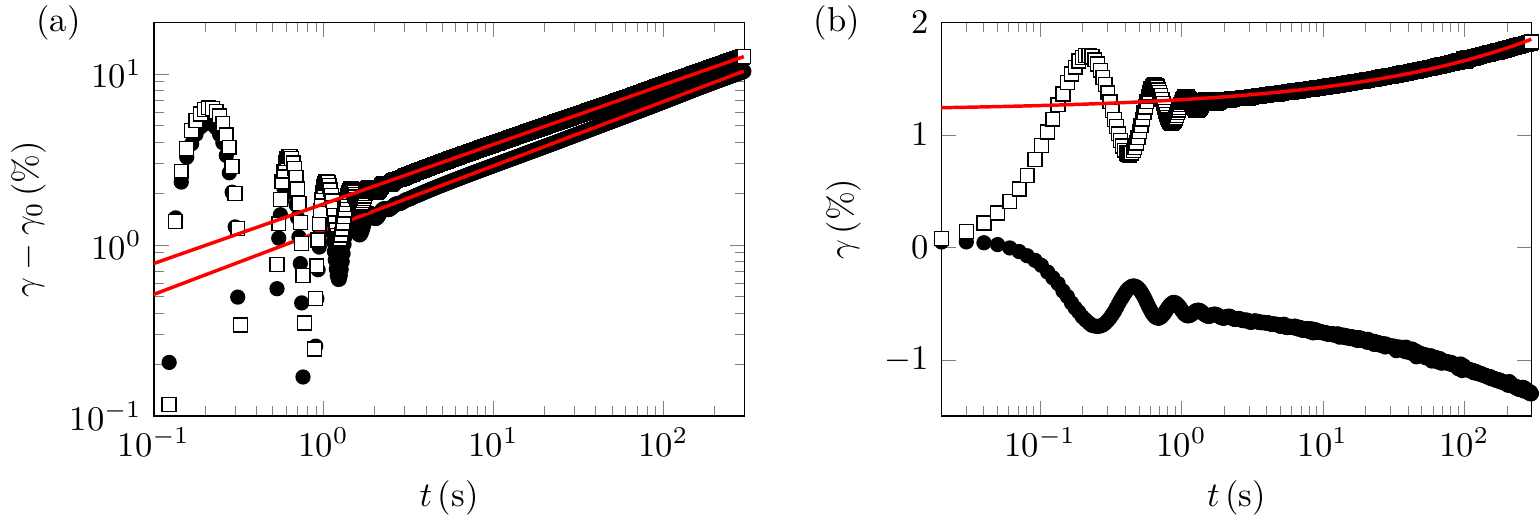}
\caption{Influence of the rest protocol on creep  in a 1~\% wt carbopol microgel. Following preshear at $\dot{\gamma}_\text{p}=\SI{100}{\per\second}$ for $t_\text{p}=\SI{10}{\second}$, a rest time of $t_\text{w}=\SI{300}{\second}$ is imposed either under a zero shear rate ($\bullet$) or under a zero shear stress ($\square$) on the same loading of the cone-and-plate geometry. (a)~Strain responses $\gamma(t) - \gamma_0$ for $\sigma=\SI{5}{Pa}$ plotted in logarithmic scales together with Andrade fits (red solid lines) yielding respectively $\gamma_0=11.4\%$ and $\alpha=0.38$ for rest under a zero shear rate and $\gamma_0=13.4\%$ and $\alpha=0.35$ for rest under a zero shear stress. (b)~Strain responses $\gamma(t)$ for $\sigma=\SI{0.4}{Pa}$ plotted in semilogarithmic scales. The red solid line is the Andrade fit in the case of rest under a zero shear stress with $\gamma_0=1.2\%$ and $\alpha=0.33$.}
\label{fig:rest1}
\end{figure*}

In order to test the influence of the rest protocol, Fig.~\ref{fig:rest1} compares creep experiments following a rest period performed either under a zero shear stress or under a zero shear rate (or more precisely under small strain oscillations with amplitude 1\% and frequency 1~Hz, which were checked to effectively correspond to a zero shear rate). As shown in Fig.~\ref{fig:rest1}(a) for  $\sigma=\SI{5}{Pa}$, the Andrade exponent is robust to a change of rest protocol but the initial deformation $\gamma_0$ is slightly smaller when rest is imposed under a zero shear rate. Moreover, when the imposed stress is decreased to $\sigma=\SI{0.4}{Pa}$, anomalous creep characterized by a decreasing strain is recovered only in the case of rest under a zero shear rate [see Fig.~\ref{fig:rest1}(b)]. In the case of rest performed under a zero shear stress, Andrade-like response persists down to the lowest imposed stresses as shown by the fit in Fig.~\ref{fig:rest1}(b).

\begin{figure*}
\includegraphics[width=0.9\textwidth]{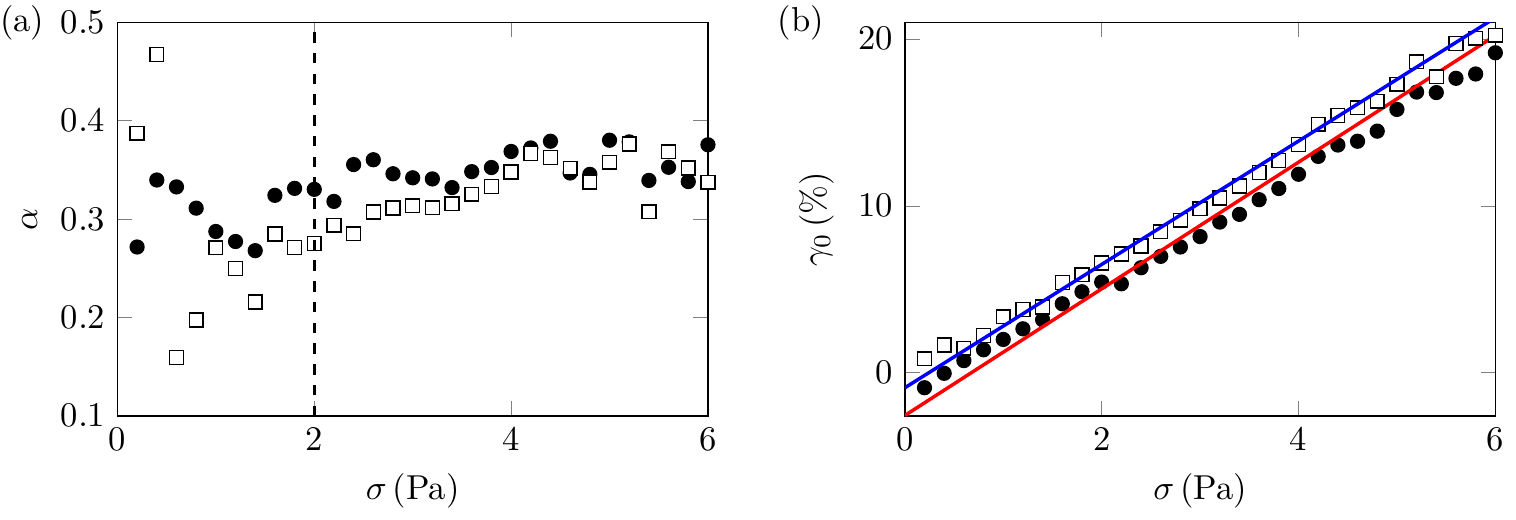}
\caption{Andrade fit parameters for the two rest protocols used in Fig.~\ref{fig:rest1} on the same loading ($\bullet$: zero shear rate, $\square$: zero shear stress): (a)~exponent $\alpha$ and (b)~initial elastic deformation $\gamma_0$, as a function of the applied stress $\sigma$. The vertical dashed line shows the typical stress of 2~Pa below which anomalous creep is observed when rest is imposed at a zero shear rate. Solid lines in (b) are linear fits $\sigma=G'_0\gamma_0+\sigma_0$ with $G'_0=\SI{26.3}{\pascal}$ and $\sigma_0=\SI{0.7}{\pascal}$ for rest under a zero shear rate (red line) and $G'_0=\SI{27.0}{\pascal}$ and $\sigma_0=\SI{0.2}{\pascal}$ for rest under a zero shear stress (blue line).}
\label{fig:rest2}
\end{figure*}

Andrade fit parameters are displayed in Fig.~\ref{fig:rest2} which confirms the good collapse of the exponents $\alpha$ independent of the rest protocol, provided the imposed stress is large enough and anomalous creep is avoided. The linear fits of $\gamma_0$ vs $\sigma$ show that the difference noted above in the initial deformation does not stem from the slope $G'_0$ which remains close to the elastic modulus of the microgel but from the intercept $\sigma_0$ which is significantly larger in the case of rest under zero shear rate (0.7~Pa) than in the case of rest under zero shear stress (0.2 ~Pa). These results indicate that imposing a zero shear stress during the rest time essentially cancels out residual stresses so that power-law creep is observed even at very low applied stresses. 

\end{document}